\documentclass[prb,aps,reprint,twocolumn,showpacs,
tightenlines,amsmath,amssymb]{revtex4-1}
\usepackage{graphicx}    
\usepackage{amsmath}     
\usepackage{amssymb}  
\usepackage{mathrsfs}
\usepackage{bm}                                          
\usepackage{colordvi}         
\usepackage{epsfig}
\newcommand{\bgreek}[1]{\mbox{\boldmath$#1$\unboldmath}}
\begin{document}

\title{
  High-field charge transport on the surface of 
  Bi$_2$Se$_3$ 
}
\author{M. Q. Weng} \email{weng@ustc.edu.cn}
\author{M. W. Wu} \email{mwwu@ustc.edu.cn}

\affiliation{Hefei National Laboratory for Physical Sciences at Microscale and
  Department of Physics, University of Science and Technology of China, Hefei,
  Anhui, 230026, China} 
 \date{\today}

 \begin{abstract}
   We present a theoretical study on the high-field charge
   transport on the surface of Bi$_2$Se$_3$ and reproduce all the main
   features of the recent experimental
   results, 
   {i.e.}, the incomplete current saturation
   and the finite residual conductance in the high applied field
   regime 
   [Costache {\it et al.}, Phys. Rev. Lett. {\bf 112}, 086601 (2014)].
   Due to the hot-electron effect, the conductance
   decreases and the current shows the tendency of the saturation with the
   increase of the applied electric field. 
   Moreover, the electric field 
   can excite carriers within the surface bands through interband
   precession and leads to a higher conductance. As a joint effect of
   the hot-electron transport and the carrier excitation, the
   conductance approaches a finite residual value in the high-field
   regime and the current saturation becomes incomplete.
   We thus demonstrate that, contrary to the conjecture in the
   literature, the 
   observed transport phenomena can be understood qualitatively 
   in the framework of surface transport alone. 
   Furthermore, if a constant bulk conductance which is insensitive to
   the field is introduced, one can obtain a good
   quantitative agreement between the theoretical results and the
   experimental data.  
 \end{abstract}

\pacs{
73.50.Fq, 
75.70.Tj, 
72.25.Rb
}

\maketitle

\section{Introduction}

The three-dimensional topological insulators (TIs) 
have attracted much attention recently due to the intriggering
fundemental physics as well as the possible application in
the TI devices.\cite{PhysRevLett.98.106803,Zhang2009,Xia2009,
PhysRevLett.103.146401,RevModPhys.82.3045,Hsieh2009,Analytis2010,
PhysRevLett.106.257004,
Chen10072009,RevModPhys.83.1057,ISI:000288917400004,
PhysRevB.78.195424,Fiete2012845}
TI has a gapped insulating bulk but
gappless conducting surface states whose low energy ones can be
described as massless Dirac fermions. 
The discovery of the strong TI materials such as Bi$_2$Se$_3$,
which has a bulk gap on the order of 300~meV, are of particular
interest since it indicates
the feasibility of the room temperature devices. 
The surface states of the strong TI materials near the Dirac points
have been experimentally measured by the spin-angle resolved photoemission
spectroscopy.\cite{Xia2009,PhysRevLett.106.257004} 
However, the signature of the surface states has yet to be
separated from the bulk ones in the transport experiments. 
Due to the relatively small gap, it is
commonly believed that the bulk states have strong 
influence on the
charge transport,\cite{Culcer2012860,PhysRevB.82.155457,
PhysRevLett.105.066802,arxiv:1310.0202,arxiv:1303.7306,
Analytis2010,PhysRevB.81.241301,
PhysRevB.81.205407,PhysRevB.81.195309,*PhysRevB.84.075316,
*PhysRevB.84.165311,*PhysRevB.85.155301}
even when the Fermi level resides inside the bulk
gap.\cite{PhysRevLett.106.196801,PhysRevB.84.165311,
PhysRevB.85.155301,PhysRevLett.112.086601}  
However, without a clear understanding of surface transport, it is
premature to distinguish the surface transport from the bulk one.

Recently, Costache {\it et al.} reported 
the experimental investigation on the charge transport on the 
surface of Bi$_2$Se$_3$ under high electric
fields.\cite{PhysRevLett.112.086601}  
In the experiment, it is observed that the current increases with the
applied voltage in low voltage regime, then shows a tendency of
saturation at the intermediate regime. However, the current
saturation is not complete. When the voltage further increases to
about 50~mV, the current rises again. For conductance, it undergoes 
slight change in the small voltage regime, then a quick 
reduction in the intermediate one, and finally saturates in the
high applied voltage regime (larger than 50~mV). 
The current saturation or the conductance reduction is attributed to
the inelastic electron-optical-phonon scattering in the surface
states. The finite saturated conductance in the high voltage regime,
however, is speculated to be the contribution from the bulk ones.
It is argued that due to band bending, the energy gap of Bi$_2$Se$_3$
is reduced to be 50~meV from the original 300~meV. Therefore, when the
applied voltage is higher than 50~mV, the carriers are excited 
from surface bands to the bulk band which has finite
conductance.\cite{PhysRevLett.112.086601} 
However, this argument is 
valid only when the transport is nearly
ballistic so that the carriers in the surface bands can have enough
energy gain to be excited to the bulk bands. However, for the diffusive
transport in the experimental
setup, the average energy gain is
estimated to be less than 4~meV when the applied voltage is 50~mV.
This indicates that the excitation of carriers from the surface bands to
the bulk one is very unlikely to have significant effect on the
transport even if the energy gap is indeed reduced to 50~meV. 
Therefore, the incomplete saturation and the finite residual
conductance are unlikely from the bulk contribution. 

Theoretically, the charge and spin transports of the surface state in
Bi$_2$Se$_3$ have been
investigated\cite{Culcer2012860,PhysRevB.82.155457,PhysRevLett.105.066802}
with most of the studies focusing on the linear transport regime.
Zhang and Wu have carried out the study on 
the hot-electron transport under strong electric field of
the surface states in Bi$_2$Se$_3$ 
by solving the kinetic spin Bloch equations
(KSBEs).\cite{PhysRevB.87.085319}  
It is shown that the mobility, and hence the conductance,
decreases with increasing applied electric field 
due to the hot-electron effect. Moreover, it is further shown that the
electric field can excite carriers on the surface from the valance
to the conduction bands through the interband precession. The excited 
carriers also contribute to the charge transport, thus lead to 
an enhanced conductance. 
Combining these two results, it is possible to understand the 
qualitative dependence of the current and the conductance on the 
applied voltage
within the framework of the surface transport {\em alone}.
In this paper, we show that 
the main features of the experimental results, {i.e.,} the
incomplete current saturation and the finite residual conductance,
can indeed be captured by the surface transport {\em alone}, 
even though the bulk does have non-neglectable contribution to the
total conductance.

This paper is organized as follows: In Sec.~\ref{model}, we set up
the model and present the KSBEs. In Sec.~\ref{results}, we show that
the main features of the experimental results, such as the incomplete
saturation of current and the finite residual conductance at high
applied field, can be captured by the surface transport
alone. Moreover, we show that one can obtain a good agreement between
the experimental results and theoretical calculation by introducing a
constant bulk conductance that does not change with the applied voltage. 
We summarize in Sec.~\ref{conclusion}

\section{Model and KSBEs}
\label{model}

The Hamiltonian for the electrons on the (001) Bi$_2$Se$_3$ surface grown
along the $z$-direction is composed of the free part $H_0$ and the
interacting part $H_I$. The free part describes the 
low energy surface states around the $\Gamma$
point and can be written in form of the
Rashba\cite{0022-3719-17-33-015} spin-orbit
coupling\cite{Zhang2009,PhysRevB.82.045122,PhysRevLett.103.266801}
\begin{equation}
  \label{eq:Hamiltonian}
  H_0=\sum_{k}v_F(\mathbf{k}\times \mathbf{\hat{z}})\cdot\bgreek{\sigma}
  c^{\dag}_{\mathbf{k}\sigma}c_{\mathbf{k}\sigma},
\end{equation}
in which, $\hbar$ is set to be 1, $v_F$
is the Fermi velocity which is chosen to be 
$2\times 10^5$~m/s,\cite{PhysRevB.82.155457} 
$c_{\mathbf{k}\sigma}(c^{\dag}_{\mathbf{k}\sigma})$ is the
annihilation (creation) operator of the electron with 
the in-plane momentum 
$\mathbf{k}=(k_x,k_y)$ and spin $\sigma=(\uparrow,\downarrow)$ and 
$\bgreek{\sigma}$ are the Pauli matrices for spin. 
In the collinear spin space spanned by the eigenstates of $\sigma_z$
($|\!\!\uparrow\rangle$ and $|\!\!\downarrow\rangle$), the surface states can
be expressed by the two helix spin states 
$|\mathbf{k}\pm\rangle={1\over \sqrt{2}}(\pm 
e^{-i\theta_{\mathbf{k}}}|\!\!\uparrow\rangle+|\!\!\downarrow\rangle)$, with
$\theta_{\mathbf{k}}$ being the polar angle of the momentum
$\mathbf{k}$. The $+$  and $-$ branches, with linear dispersion
$\varepsilon_{\mathbf{k}\pm}=\pm v_Fk$, correspond to the
conduction and valance bands of the surface states, respectively. 
The interacting part $H_I$ describes the electron-impurity
scattering, electron-phonon coupling and electron-electron Coulomb
interaction. It can be written as 
\begin{eqnarray}
  &&H_I=\sum_{\mathbf{q},\mathbf{k}\sigma}
  v_{\mathbf{q}}\rho_I(\mathbf{q})
  c^{\dag}_{\mathbf{k}+\mathbf{q}\sigma}
  c^{\dag}_{\mathbf{k}\sigma} \nonumber \\ 
  &&+
  \sum_{\lambda\mathbf{q}\Omega,\mathbf{k}\sigma}
  M_{\lambda}(\mathbf{q},\Omega)
  \phi_{\lambda}(\mathbf{q},\Omega)
  c^{\dag}_{\mathbf{k}+\mathbf{q}\sigma}
  c_{\mathbf{k}\sigma} \nonumber \\ && 
  +  
  \sum_{\mathbf{q}\mathbf{k}'\sigma'\mathbf{k}\sigma}
  v_{\mathbf{q}}
  c^{\dag}_{\mathbf{k}'-\mathbf{q}\sigma'}
  c^{\dag}_{\mathbf{k}+\mathbf{q}\sigma}  
  c_{\mathbf{k}\sigma}  
  c_{\mathbf{k}\sigma'}. 
  \label{eq:HI}
\end{eqnarray}
Here $v_{\mathbf{q}}=e^2/(2\varepsilon_0\kappa_0 q)$ with $e$,
$\varepsilon_0$ and $\kappa_0$ standing for the elementary charge,
permittivity, and the dielectric constant of Bi$_2$Se$_3$,
respectively.  
$\kappa=100$.\cite{PSSB:PSSB2220840226,PhysRevB.81.241301,PhysRevB.87.085319} 
$\rho_I(\mathbf{q})=\sum_{i=1}^{N_i}e^{i\mathbf{q}\cdot
  \mathbf{R}_i}$, where $\mathbf{R}_i$ is the position of $i$-th
impurity and  $N_i$ is the impurity density. 
$\phi_{\lambda}(\mathbf{q},\Omega)=b_{\lambda}(\mathbf{q},\Omega)+
b^{\dag}_{\lambda}(-\mathbf{q},\Omega)$, with
$b_{\lambda}(\mathbf{q},\Omega)$
[$b^{\dag}_{\lambda}(\mathbf{q},\Omega)$] being the annihilation
(creation) operators of the phonon with branch $\lambda$, momentum
$\mathbf{q}$ and energy $\Omega$. 
For the electron-phonon coupling, we include contributions
from the surface optical phonon, longitudinal and transverse acoustic
phonons.\cite{PhysRevB.83.245322,*PhysRevB.85.035441} 
The matrix elements for the electron-surface optical phonon coupling
read
\begin{equation}
  \label{eq:optphonon}
  M_{\mathrm{op}}(\mathbf{q},\Omega=\omega_o)=(\lambda_1+\lambda_2
  q)/\sqrt{2M{\cal A}\omega_o}, 
\end{equation}
where $M$ is the ion mass, ${\cal A}$ is the primitive cell area
[$1/(M{\cal A})=4\times 10^{-3}$meV], 
$\lambda_1=5$~eV~nm, $\lambda_2=1.6$~eV~nm$^2$ and
$\omega_o=8$~meV is the optical phonon
energy.\cite{Zhang2009,PhysRevLett.107.186102}  
For the longitudinal and transverse acoustic phonon, 
the matrix elements of the electron-phonon coupling read
\begin{equation}
  \label{eq:longphonon}
  M_{L}(\mathbf{q},\Omega)=-\alpha {(\Omega/v_l)^2\over
    \sqrt{2\rho_M\Omega}} {(q^2-k_t^2)^2-4q^2k_lk_t\over
    (q^2-k_t^2)^2+4q^2k_lk_t}
  \Theta(\Omega-v_lq),
\end{equation}
\begin{equation}
  \label{eq:transphonon}
  M_{T}(\mathbf{q},\Omega)=-\alpha {(\Omega/v_l)^2\over
    \sqrt{2\rho_M\Omega}} {4q(q^2-k_t^2)\sqrt{k_lk_t}\over
    (q^2-k_t^2)^2+4q^2k_lk_t}
  \Theta(\Omega-v_tq),
\end{equation}
respectively.\cite{PhysRevB.83.245322,PhysRevB.85.035441}
In the above equations, 
$k_{l,t}=\sqrt{(\Omega/v_{l,t})^2-q^2}$, $v_{l(t)}=2900(1700)$~m/s is the
longitudinal (transverse) sound
velocity,\cite{PhysRev.185.1046,PSSB:PSSB2220840226}  
$\rho_M=7860$~kg/m$^3$ is the mass density\cite{Wiese196013,PhysRevB.76.045302} 
and $\alpha=70$~eV.\cite{PhysRevB.85.035441}

By using the nonequilibrium Green function method,\cite{haugjauho}
we construct the KSBEs for spatially uniform system 
as follows\cite{Wu201061,cheng:083704,PhysRevB.87.085319} 
\begin{eqnarray}
  &&\partial_t\rho_{\mathbf{k}}(t)+i[v_Fk\sigma_z,\rho_{\mathbf{k}}(t)]
  -eE\partial_{k_x}\rho_{\mathbf{k}}(t) \nonumber \\ 
  &&\hspace{1pc}
  -eE[U^{\dag}_{\mathbf{k}}\partial_{k_x}U_{\mathbf{k}},
  \rho_{\mathbf{k}}(t)]
  +\partial_t\rho_{\mathbf{k}}(t)|_{\mathtt{scat}}=0. 
  \label{eq:KSBE}
\end{eqnarray}
Here $\rho_{\mathbf{k}}(t)$ is the density matrix for electrons with
momentum $\mathbf{k}$ in the helix spin space. The diagonal elements
of $\rho_{\mathbf{k}}(t)$,
$\rho_{\mathbf{k}++/--}(t)=f_{\mathbf{k}+/-}(t)$, represent the
electron distribution functions in the conduction and valance bands,
respectively, while the off-diagonal terms stand for the interband
coherence. The KSBEs include the coherent term (the second term in the
left-hand-side of the equation), the 
acceleration of the electron under the electric field $E$ (the third
term), the interband precession induced by the field (the
fourth term) as well as the scattering term (the fifth term). 
The electric-field-induced interband precession originates from the
spin mixing in the conduction and valance bands. Similar effect also
exists in the graphene where the pseudospins are
mixed.\cite{PhysRevB.79.165432} This term is
usually ignored in the previous studies on low field
transport.\cite{PhysRevB.79.165432} 
However,
it is shown that the electric-field-induced interband precession has
profound influence on the transport properties under high electric
field as it leads to a strong excitation of the carriers from the
surface valance to the conduction bands.\cite{PhysRevB.87.085319} 
The expressions for the scattering term 
can be found in Ref.~[\onlinecite{PhysRevB.87.085319}], 
which includes the contribution from electron-electron Coulomb
interaction, electron-impurity scattering and electron-phonon
coupling. 

By numerically solving the KSBEs for spatially uniform
system, one obtains the charge current density
$J$ for the applied electric field $E$. 
We apply our results to a sample with length $L=410$~nm and
width $W=300$~nm,
the same length and width as those of the sample D2 
in Ref.~[\onlinecite{PhysRevLett.112.086601}], 
in order to obtain the quantities
such as the applied voltage $V=EL$ and 
the conductance $G=I/V$ with the current $I=2JW$.
Here the prefactor 2
is from the fact that each sample contains two surfaces.


\section{Numerical Results}
\label{results}

\begin{figure}[!htbp]
  \centering
  \epsfig{file=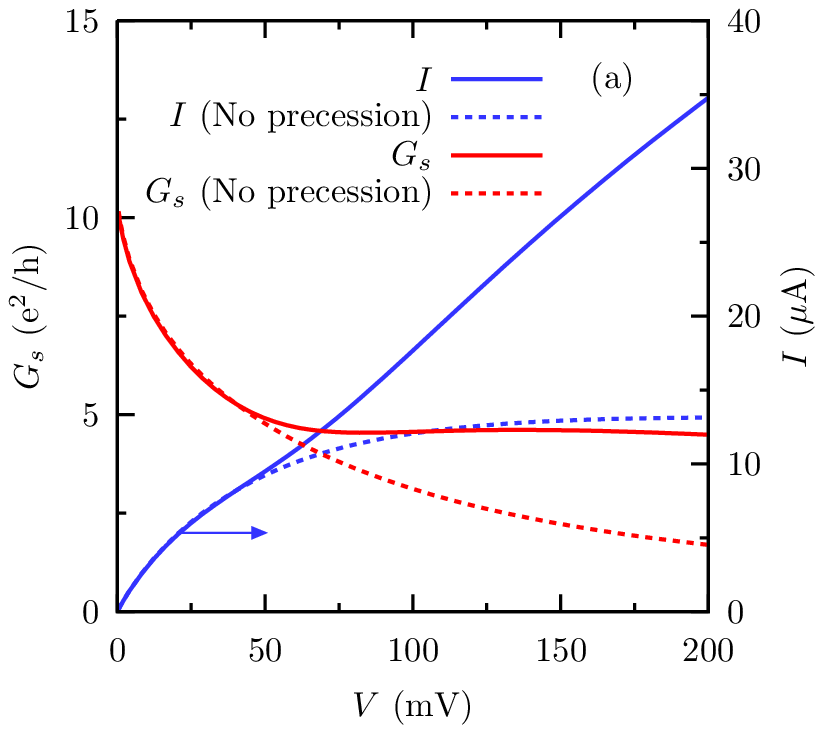,width=8cm}
  \epsfig{file=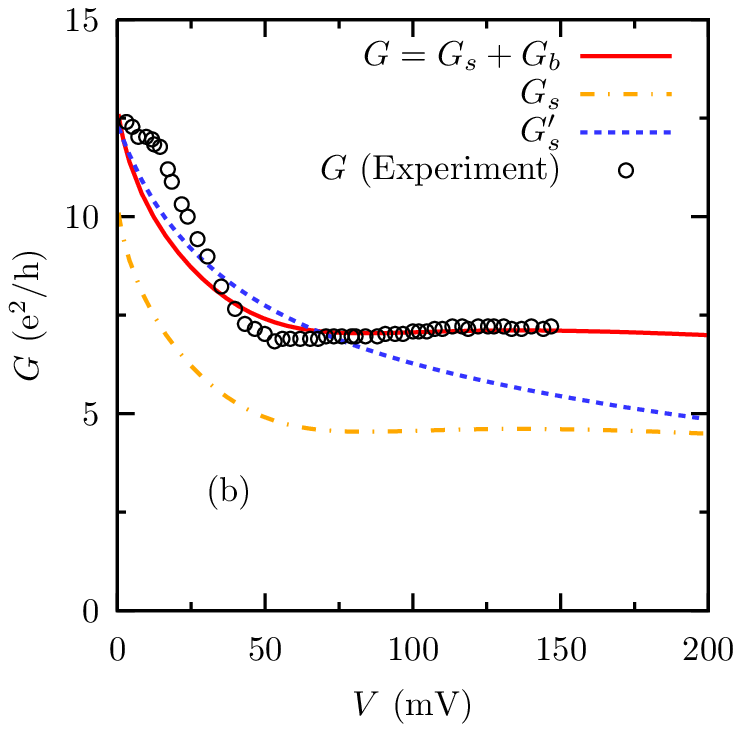,width=8cm}
  \caption{(Color online) (a) 
    Current (blue curves) and conductance
    (red curves) as function of the applied voltage 
    at temperature $T=4.2$~K for surfaces with $N_e=1.5\times
    10^{11}$~cm$^{-2}$ and $N_i=6\times 10^{11}$~cm$^{-2}$. 
    The solid and the dashed curves are the results with and without
    the interband precession. 
    Note that the scale of the current is on the right hand side of
    the frame.
    (b) The conductance as function of the applied voltage 
    at temperature $T=4.2$~K with different fitting parameters:
    The blue dashed curve is the surface conductance $G'_s$ for the
    surfaces with 
    $N_e=9.36\times 10^{11}$~cm$^{-2}$ and $N_i=2.1\times
    10^{12}$~cm$^{-2}$; 
    The dark yellow dash-dotted curve is the surface conductance $G_s$ for
    the surfaces with 
    $N_e=1.5\times 10^{11}$~cm$^{-2}$ and $N_i=6\times
    10^{11}$~cm$^{-2}$, which is exactly the red solid curve in (a). 
    The red solid curve is
    the total conductance for the surfaces with 
    $N_e=1.5\times 10^{11}$~cm$^{-2}$ and $N_i=6\times
    10^{11}$~cm$^{-2}$
    plus a bulk with a constant conductance 
    $G_b=2.5$~e$^2$/h. 
    The circles are the experimental data from
    Ref.~[\onlinecite{PhysRevLett.112.086601}]
    for sample D2 under the gate
    voltage of $-120$~V.
  }
  \label{fig:conductance}
\end{figure}

In Fig.~\ref{fig:conductance}(a), we plot the current $I$
as a function of the applied voltage $V$ at 4.2~K
for the surface states with 
electron density $N_e=1.5\times 10^{11}$~cm$^{-2}$ and 
impurity density $N_i=6\times 10^{11}$~cm$^{-2}$. 
One can see that our theoretical results capture all the main
experimental results qualitatively. Namely, the current $I$ increases
linearly with the increase of the 
voltage $V$ when the voltage is small; Then the
current shows a tendency of saturation at intermediate voltage but
increases 
again when the voltage rises higher than $50$~mV. This incomplete
saturation can also be seen in the voltage dependence of the
conductance $G$, also shown in Fig.~\ref{fig:conductance}(a). One finds
from the figure that, the conductance decreases as the voltage increases
then saturates to a finite constant at higher voltage. 

The decrease of the
conductance is a result of the hot-electron effect at high electric
field.\cite{PhysRevB.69.245320,PhysRevB.87.085319} 
Due to the driving of the electric field, the 
temperature of the electrons raises well above the lattice one
when the field is high enough.
As a result, the electron-phonon scattering is profoundly enhanced,
consequently the mobility $\mu$ decreases. Therefore, the conductance
$G=N_ee\mu EW/V$ decreases with the 
increase of the voltage. When the field is
strong enough, the mobility under electric field $E$
is roughly inversely proportional to the
electric field $1/\mu\simeq 1/\mu_0+\gamma(N_e) E$.
Here $\mu_0$ is the linear mobility and 
$\gamma(N_e)$ is a constant that weakly decreases when carrier
concentration $N_e$ rises.\cite{PhysRevB.87.085319} 
One can therefore estimate that the current should saturates to
$I_s=N_eeW\gamma(N_e)$ at high voltage. This current saturation can be
crudely captured by the so called steady-state population model with
instantaneous phonon emission.\cite{PhysRevLett.103.076601,PhysRevLett.84.2941} 

If the saturation is complete as predicted by the steady-state
population model, the current should remain constant and
the conductance should approach zero under high voltage. 
However, our computation shows that at higher applied electric field,
the current saturation is not complete. 
When the voltage further increases 
the current again rises linearly with the voltage, and the
corresponding conductance saturates to a finite value instead of
zero. This incomplete current saturation and the finite residual
conductance in the high-field regime
is due to the excitation of carriers from
the surface valence band to the surface conduction one
via the interband precession (the
fourth term in KSBEs),
first proposed by Zhang and Wu.\cite{PhysRevB.87.085319} 
In the low field
regime, the excitation is proportional to $E^2$. For sample with a high
background carrier density, the number of the excited carriers is too
small to have an observable effect on transport properties. 
In the intermediate regime where the mobility is reduced due to the
hot-electron effect but the carrier excitation is not yet strong
enough, the current shows tendency of saturation as
the mobility reduces.
At higher field regime, when more carriers are excited, 
the current increases again with the increases of the
field. Under strong field, the excited carrier density is proportional
to the electric field $E$,
and the total number of the carriers becomes 
\begin{equation}
  \label{eq:NeE}
  N_e(E)\simeq N_e+\beta E,
\end{equation}
with $\beta$ being a constant. 
Combining with the fact that the mobility is inversely proportional to
$E$ in this regime, one finds that the conductance 
\begin{equation}
  \label{eq:ConductanceE}
  G\simeq [N_e+\beta E]e\mu_0 W/[(1+\gamma(N_e)\mu_0 E)L]. 
\end{equation}
It is therefore understood that under strong applied
electric field/voltage when the number of the excited carriers exceeds
the background one, 
the conductance becomes a finite constant 
\begin{equation}
  \label{eq:ResidualConductance}
  G_r\simeq \beta
  eW/[\gamma(N_e)L].
\end{equation}
The saturated field $E_s$ is determined by the equation
$\beta E_s=N_e$, {i.e.}, when the number of the excited carriers
becomes the same  as the background one. One can then write down 
the saturated voltage $V_s$ as
\begin{equation}
  \label{eq:SaturatedVoltage}
  V_s=E_sL=N_eL/\beta,
\end{equation}
which is proportional to the background surface charge density.

If there is no such excitation, the current saturation at high applied
field/voltage regime is complete and the conduction
approaches zero. This is verified by our numerical solution,
also shown in Fig.~\ref{fig:conductance}, with the
interband precession term in the KSBEs 
artificially removed.  

To check the quantitative agreement between the experimental results
and the theoretical ones, we plot our numerical fittings to
the experimental data from Ref.~[\onlinecite{PhysRevLett.112.086601}]
for sample D2 under a gate voltage of  $-120$~V by using different
parameters in Fig.~\ref{fig:conductance}(b). 
The electron density per surface is estimated to be $9.4\times 
10^{11}$~cm$^{-2}$ in the experiment at such a gate voltage.
If all the electrons occupy the surface states and the
bulk does not have any contribution to the charge transport, then the 
total conductance $G$ is the same as the surface one. 
Under this assumption, the surface conductance from our computation
($G_s'$), shown in
Fig.~\ref{fig:conductance}(b) as dashed curve, fits 
reasonably well with the 
experimental data in the low field regime when the
impurity density is $2.1\times 10^{12}$~cm$^{-2}$. 
In the high field regime,
however, the theoretical results deviate from
the experimental ones. The theoretical saturated voltage in this case
is about 200~mV, much larger than 50~mV from the experiment, and the
residual conductance is about 5$e^2/h$, 30~\% smaller than the
experimental one.

According to Eq.~(\ref{eq:SaturatedVoltage}), to have a smaller
saturated voltage, the background surface charge density should be
smaller. The overall quantitative fitting can indeed be improved if
some of the 
electrons are assumed to be populated in the bulk impurity 
bands.\cite{PhysRevLett.106.196801,PhysRevLett.112.086601} 
In this case, the total conductance $G$ is the summation of surface
conductance $G_s$ and the bulk one $G_b$. 
Since the conductance of the bulk impurity bands is determined by the
disorder, $G_b$ can be regarded as a constant that does not
variate with the applied voltage in the observed
regime.\cite{PhysRevLett.106.196801,PhysRevLett.112.086601}   
In Fig.~\ref{fig:conductance}(b) we further show
the total conductance 
$G$ for the surfaces with $N_e=1.5\times 10^{11}$~cm$^{-2}$ and
$N_i=6\times 10^{11}$~cm$^{-2}$ plus a bulk contribution with a constant
conductance $G_b=2.5$$e^2/h$. One finds that our numerical results
are in good agreement with the 
experimental data for both low and high
applied voltages.
It is noted that we do not consider the excitation of carriers from
the surface states to the bulk ones
in our calculation. The charge
densities on the surface and in the bulk remain constant and do not
change with the applied field/voltage. Since the bulk charge density
and the conductance $G_b$ are constant, all the variations in the
transport properties 
are caused solely by the carriers on the surface.
Our results suggest that the bulk indeed has a non-negligable
contribution to the total conductance. However, the main 
transport phenomena, such as the incomplete current saturation and
the finite residual conductance under high field, can be understood
within the framework of surface transport alone without the influence
of the bulk part. 

In the works of Checkelsky {\it et al.}\cite{PhysRevLett.106.196801} and 
Costache {\it et al.},\cite{PhysRevLett.112.086601}
the incomplete current saturation and the finite residual conductance
were speculated to be associated with
the excitation of the carrier from surface 
states to the bulk ones
which have finite constant conductance.  
It is argued that in the experimental
setup, the energy gap of
Bi$_2$Se$_3$ can be reduced from 300~meV to about 50~meV due to the
band bending, thus 
enabling the carriers 
on the surface to be excited to the bulk
with an applied voltage of 50~mV. 
For the argument to be valid, the transport must be nearly ballistic so
that the carriers in the surface states can gain enough energy to jump
over the energy gap. However, in the experimental setup the mean free
path is less than 30~nm. It means that the average energy gain by the
carriers is less than 4~meV for an applied voltage of 50~mV over a
sample with a length of 410~nm. This energy gain is too small to excite
the carriers from the surface 
states to the bulk ones even if the
energy gap is indeed reduced to 50~meV. Therefore, the excitation from
the surface states to the bulk ones is very unlikely to be the main
reason for the incomplete current saturation and the finite residual
conductance at high voltage. From our theoretical results, the more
likely reason for these 
high-field/voltage
transport phenomena is the
joint effects of the hot-electron transport and the
excitation within the surface bands. 

\section{Conclusion}
\label{conclusion}

In conclusion, we have studied the charge transport on the surface of
Bi$_2$Se$_3$ by numerically solving the KSBEs and reproduce the main 
qualitative features of the experimental results, {i.e.}, the
incomplete current saturation and the finite residual conductance in
high applied voltage regime, without introducing any bulk
contribution. 
Due to the hot-electron effect, the electron-phonon scattering is
enhanced and leads to a reduced mobility, inversely
proportional to the applied field at high field regime. As a result,
the current shows tendency of saturation.
On the other hand, the applied electric field can
excite carriers from the surface valance band to the surface
conduction one due to the interband preccession. This leads to the
increase of the current carrying carriers
and thus the increase of
current. Under high applied voltage, the conductance approaches a
finite residual value as the number of the excited
carriers, being proportional to the electric field, exceeds the
background one. Moreover, the theoretical results agree quantitatively
well with the experimental data if a constant bulk conductance, which does not
change with the applied voltage in the experimental measured regime,
is introduced. This suggests   
that even though the bulk has a non-neglectable contribution to the
the total conductance, the main transport phenomena, such as the
incomplete current saturation and the finite residual conductance at
high applied voltage/field, can be well understood in the framework of
surface transport alone.

\begin{acknowledgments}
This work was supported by the National Natural Science
Foundation of China under Grant No. 11334014,
the National Basic Research Program of China under
Grant No. 2012CB922002, and the Strategic Priority Research
Program of the Chinese Academy of Sciences under
Grant No. XDB01000000
\end{acknowledgments}


%

\end{document}